\documentclass[aps,pra,reprint,superscriptaddress,showpacs,amsmath,amssymb]{revtex4-1}

\usepackage{amsmath,amssymb,amsfonts}

\usepackage{graphicx}
\usepackage{epstopdf}
\usepackage{xcolor}
\DeclareGraphicsExtensions{.eps}

\definecolor{darkblue}{rgb}{0, 0, 0.8}
\usepackage[colorlinks=true, breaklinks=true, linkcolor=darkblue, citecolor=darkblue, urlcolor=darkblue]{hyperref}
\newcommand{\doilink}[2]{\href{http://dx.doi.org/#1}{#2}}

\begin{document}
\title{Polaritonic modes in a dense cloud of cold atoms}
\date{\today}
\author{N.J. Schilder}
\affiliation{Laboratoire Charles Fabry, Institut d'Optique Graduate School, CNRS, Universit\'e Paris-Saclay, 91127 Palaiseau Cedex, France}
\author{C. Sauvan}
\affiliation{Laboratoire Charles Fabry, Institut d'Optique Graduate School, CNRS, Universit\'e Paris-Saclay, 91127 Palaiseau Cedex, France}
\author{J.-P. Hugonin}
\affiliation{Laboratoire Charles Fabry, Institut d'Optique Graduate School, CNRS, Universit\'e Paris-Saclay, 91127 Palaiseau Cedex, France}
\author{S. Jennewein}
\affiliation{Laboratoire Charles Fabry, Institut d'Optique Graduate School, CNRS, Universit\'e Paris-Saclay, 91127 Palaiseau Cedex, France}
\author{Y.R.P. Sortais}
\affiliation{Laboratoire Charles Fabry, Institut d'Optique Graduate School, CNRS, Universit\'e Paris-Saclay, 91127 Palaiseau Cedex, France}
\author{A. Browaeys}
\affiliation{Laboratoire Charles Fabry, Institut d'Optique Graduate School, CNRS, Universit\'e Paris-Saclay, 91127 Palaiseau Cedex, France}
\author{J.-J. Greffet}
\affiliation{Laboratoire Charles Fabry, Institut d'Optique Graduate School, CNRS, Universit\'e Paris-Saclay, 91127 Palaiseau Cedex, France}

\begin{abstract}
We analyze resonant light scattering by a cloud of cold atoms with randomly distributed positions in a regime where near-field interactions between scatterers cannot be neglected. Using a microscopic approach we calculate numerically the collective eigenmodes of the cloud for many realizations. It is found that there always exists a small number of polaritonic modes. Using a macroscopic approach, we show that the atomic cloud is equivalent to a dielectric particle with an effective permittivity, and that there is a one-to-one correspondence between the microscopic polaritonic modes and the modes of the dielectric particle.

\end{abstract}
\maketitle
Light scattering by an ensemble of particles is usually viewed as a multiple scattering process involving a sequence of single scattering events. Each scatterer is considered to be in the far-field of the others so that a scattering matrix approach can be used~\cite{Bohren}.  When the density of the system increases, correlations and near-field interactions (the latter is often called recurrent scattering or, for atoms, resonant van der Waals interaction) between scatterers start playing a role. These effects and also coherence effects can be included using a coherent theory of multiple scattering~\cite{Frisch,Sheng,Lagendijk,Greffet}. Within this framework, it is possible to show that the electric field averaged over an ensemble of realizations of the system obeys a propagation equation in a homogeneous system with an effective dielectric permittivity. Many standard models are available to derive the effective permittivity in the dense regime~\cite{Maxwell,Bruggeman,Sheng,Sentenac}. The regime where the recurrent scattering becomes significant remains largely unexplored experimentally.

The physics of multiple scattering of light has been revisited in the context of resonant atomic gases in the last fifteen years. Backscattering enhancement has been measured in the low intensity regime where a classical description of scattering is valid~\cite{Labeyrie,Wilkowski}. Cold atoms allow changing the scattering regime by detuning the laser: from optically thin to optically thick media~\cite{Labeyrie2} and from dilute to dense media~\cite{Antoine}. In the dense regime, novel effects are expected as pointed out by Morice \textit{et al.}~\cite{Morice}. The recurrent scattering has been studied theoretically in Refs.~\cite{Janne,Javanainen}. These works predict that recurrent scattering becomes significant when the dipole-dipole interaction strength parameter $\rho/k^3$ is larger than unity, where $\rho$ is the number of atoms per unit volume and $\lambda_0=2\pi/k$ is the atomic resonance wavelength. It is interesting to note that this condition can be cast in the form $kl<1$ where $l\sim k^2/\rho$ is the elastic mean free path on resonance. This condition is thus similar to the so-called Ioffe-Regel criterion of Anderson localization~\cite{Sheng}. In this density regime, it has been shown that the effective dielectric permittivity can be negative~\cite{Havey2011}. The decrease of the elastic mean free path due to recurrent scattering has been discussed in Refs.~\cite{Sokolov,Havey3,Chomaz}.
Another interesting aspect of scattering by a cold-atom system is the presence of collective effects. Recently, single-photon superradiance has been studied both theoretically~\cite{Akkermans} and experimentally~\cite{Felinto}. Bienaim\'{e} \textit{et al.} described light scattering quantum optically for a finite size system and found subradiant and superradiant states~\cite{Kaiser}. They predicted that superradiance gives rise to directional light scattering. It has also been reported that spontaneous emission is dominated by superradiant modes and that the emission spectrum displays both negative and positive Lamb shifts in the presence of resonant van der Waals interactions~\cite{Evers}. Sokolov \textit{et al.}~\cite{Havey1} analyzed the response of an ensemble of cold resonant scatterers  and they compared a modal microscopic description with a macroscopic description based on the concept of effective permittivity; both approaches are in good agreement for $\rho \lesssim 0.1k^3$. Their results show the importance of polaritonic modes, which are collective excitations of dipoles coupled to the electromagnetic field~\cite{Hopfield1958}, a concept largely used in the description of the optical response of ordered condensed matter systems.

\begin{figure}
\includegraphics[width=\linewidth]{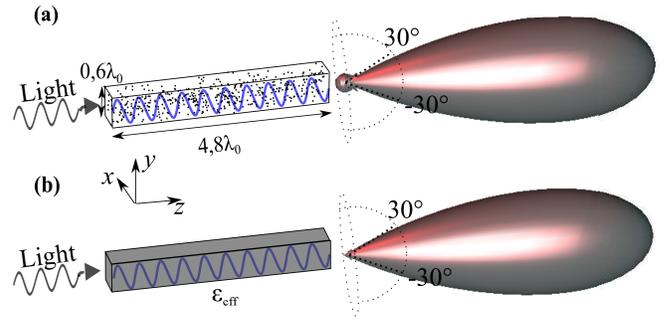}
\caption{(a) System under study : $N=450$ atoms uniformly distributed in a rectangular box. The atomic density is $\rho=k^3$. The incoming light excites the collective eigenmodes of the system (visualized by the blue sinusoid), which gives rise to a scattering pattern in, essentially, the forward direction. Here, we have plotted the modulus square of the electric field radiated in the far field at resonance, averaged over 1500 realizations, $\langle|\textbf{E}|^2\rangle$.(b) Homogeneous medium described by a dielectric permittivity $\varepsilon_{\text{eff}}$ exhibiting optical resonances. The scattering pattern, $|\textbf{E}|^2$, is found to be similar to (a) except for the diffuse light, which is null in (b).}
\label{fig:Sample}
\end{figure}

In this work, we study dense ($\rho \gtrsim k^3$) and disordered systems using, on one hand, a microscopic model including collective effects and recurrent scattering, and on the other hand a macroscopic model based on a homogeneous system with an effective permittivity. We find that one can exhibit an effective permittivity even for dense systems and that the microscopic polaritonic modes can be identified with the macroscopic modes of the homogeneous system. To demonstrate this identity, we study theoretically the microscopic and the macroscopic modes of an atomic system consisting of $N$ resonant dipoles~\cite{Antoine}. The atoms occupy a small volume with dimensions comparable to the resonant wavelength $\lambda_0$ [see Fig.~\ref{fig:Sample}a]. The dipoles being cold atoms in vacuum, they experience no non-radiative losses and a negligible Doppler effect with respect to their radiative linewidth $\Gamma_0$. In the numerical simulations presented below, we have taken rubidium-$87$ atoms, $\lambda_0=780$\,nm, and $\Gamma_0=2\pi\times(6\,{\rm MHz})$.

The formalism developed in this work is not limited to this system only, but is also applicable to condensed matter systems such as dense systems of quantum dots or organic molecules. Spatial coherence, superradiance and strong coupling with surface plasmons have been observed in these systems~\cite{Bellessa,Ebbesen,Torma,Bellessa2,Torma2,Wiersma90}.

The paper is organized as follows. First, using a microscopic model we study the collective eigenmodes of the system and show that some of them are polaritonic. We show that these modes correspond to situations where all dipoles oscillate. Also, the spatial structure of these modes does not depend on the particular positions of the scatterers. We then analyze light scattering by the system using this model and compare it with light scattering by a homogeneous system with an effective permittivity, as sketched in Fig.~\ref{fig:Sample}.
We show that polaritonic modes can be identified with the macroscopic optical modes of that system. We also discuss the origin of losses in the effective permittivity.

We consider a cloud of atoms uniformly distributed in a rectangular box (see Fig.~\ref{fig:Sample}a). The number of atoms, $N=450$, and the dimensions of the box, $4.8\lambda_0 \times 0.6\lambda_0 \times 0.6\lambda_0$, correspond to typical experimental conditions obtained by laser cooling and trapping techniques in wavelength-size optical dipole traps~\cite{Antoine}. With such parameters, $\rho/k^3\sim 1$ so that recurrent scattering starts playing a significant role.
The cloud is investigated in the weak-excitation limit, where its optical properties can be described by classical optics~\cite{Evers,Scully}. The system being dense, we treat the electromagnetic field vectorially to properly take into account the $1/r^3$-dependence of the near-field interactions ($r$ is the interatomic distance). The atoms are modeled as point-like and identical scatterers characterized by an isotropic electric polarizability matrix $\bar{\bar{\alpha}}(\omega) = \alpha(\omega)\openone$, where $\alpha(\omega)=(3\pi\Gamma_0/k^3)/(\omega_0-\omega-i\Gamma_0/2)$, with $\omega_0 = ck$ the transition frequency and $c$ the speed of light in vacuum. This polarizability model corresponds to a classical $J=0\rightarrow J=1$ atom with three transitions. It assumes elastic scattering events and therefore no non-radiative decay channels.

The microscopic eigenmodes of the atomic cloud are obtained by searching for a self-consistent solution of the coupled-dipole equations in the absence of a driving electric field. The stationary solution is one where each oscillating dipole drives the others. The induced dipoles $\textbf{p}_i$ can be written as $\textbf{p}_i=\varepsilon_0\bar{\bar{\alpha}}(\omega)\sum\limits_{j\neq i}\textbf{E}_{j\rightarrow i}(\textbf{r}_i)$, where $\textbf{E}_{j\rightarrow i}(\textbf{r}_i)=\mu_0\omega^2\bar{\bar{G}}(\textbf{r}_j,\textbf{r}_i;\omega)\textbf{p}_j$ is the electric field at the position of scatterer $i$ created by the induced dipole of scatterer $j$. Here, $\bar{\bar{G}}(\textbf{r}_j,\textbf{r}_i;\omega)$ is the vacuum Green tensor~\cite{Novotny2006}. The dipoles are thus coupled through the following linear system of equations:
\begin{equation}
\textbf{p}_i-\frac{\omega^2}{c^2}\alpha(\omega)\sum\limits_{j\neq i}\bar{\bar{G}}(\textbf{r}_j,\textbf{r}_i,\omega)\textbf{p}_j=0.
\label{eq4}
\end{equation}
The eigenmodes supported by the cloud correspond to the solutions of Eq.~\ref{eq4}. Each eigenmode has a complex frequency which cancels the system determinant. Assuming $\omega^2\bar{\bar{G}}(\textbf{r}_j,\textbf{r}_i,\omega) \approx \omega_0^2\bar{\bar{G}}(\textbf{r}_j,\textbf{r}_i,\omega_0)$ turns this problem into an eigenvalue problem, of the form $\bar{\bar{A}}(\omega_0)\textbf{P}_\beta = (\omega_\beta -i \frac{\Gamma_\beta}{2})\textbf{P}_\beta$. Because of the narrow atomic linewidth and of the non-resonant character of the vacuum Green tensor, this is an accurate assumption.

This eigenvalue problem has $3N$ eigenvalues. The corresponding eigenvectors $\textbf{P}_\beta$ are composed of the vector components of all $N$ dipole moments, $\textbf{P}_\beta=[p_{1x}^\beta,p_{1y}^\beta,p_{1z}^\beta,\cdots p_{Nx}^\beta,p_{Ny}^\beta,p_{Nz}^\beta]^\intercal$. They are normalized such that $|\textbf{P}_\beta|^2=\sum_j |\textbf{p}_j^\beta|^2=1$.

\begin{figure}
\includegraphics[width=1\linewidth]{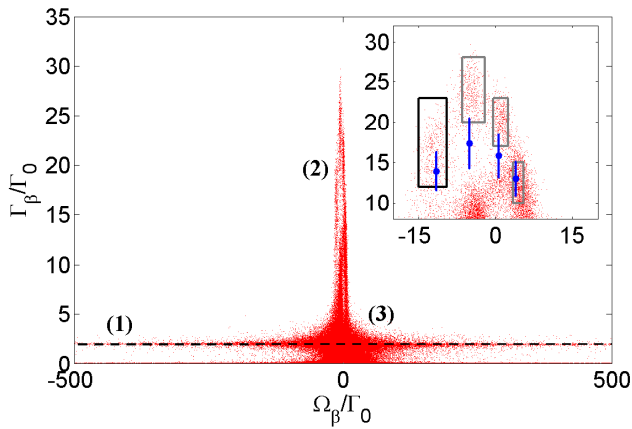}
\caption{Distribution of the $405,000$ eigenvalues obtained from $300$ realizations for $\rho/k^3=1$. The black horizontal dashed line indicates a collective decay rate $\Gamma=2\Gamma_0$. Inset: enlarged view of the main panel, showing four patches of polaritonic modes (grouped in black and gray rectangles).
Blue dots: macroscopic modes of the equivalent homogeneous cloud. Error bars are from the Lorentzian fit of the effective permittivity (see text).
}
\label{fig:Eigenvalues}
\end{figure}

\begin{figure}[b!]
\includegraphics[width=1\linewidth]{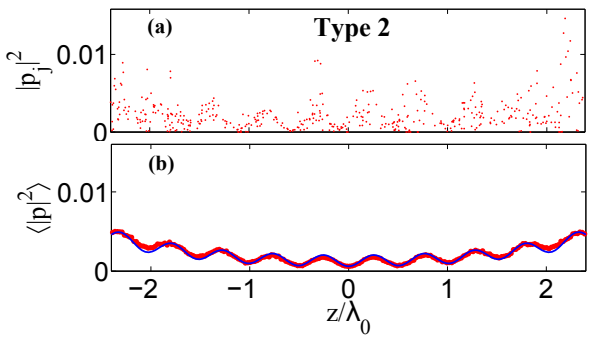}
\vspace{-1.5cm}
\caption{(a) Spatial structure of a typical microscopic eigenmode of type 2 for $\rho/k^3=1$. The dipole moment squared $|\textbf{p}_j|^2$ of each atom $j$ has been plotted, the atoms being sorted by their $z$-coordinate. (b) Average dipole moments squared $\langle|\textbf{p}|^2\rangle$ (red dots). The ensemble-average is performed over the polaritonic modes inside the black rectangle in inset of Fig.~\ref{fig:Eigenvalues}. The macroscopic mode of the homogeneous cloud (blue solid line) coincides very well with the average polaritonic mode, taking into account a small offset due to the fluctuations of the dipoles amplitudes from one realization to another~\cite{SM}.}\label{fig:Excitation}
\end{figure}

Figure~\ref{fig:Eigenvalues} shows all $405,000$ eigenfrequencies in the complex plane obtained from $300$ realizations of the cloud.
We observe three distinct families of modes, each defined in terms of their collective linewidth $\Gamma_{\beta}$ and their collective frequency shift $\Omega_{\beta}=\omega_{\beta}-\omega_0$: (1) $\Gamma_{\beta}\in\{0,2\Gamma_0\}$ and $|\Omega_{\beta}|\gg\Gamma_0$, (2) $\Gamma_{\beta}\gg\Gamma_0$ and (3) others. Also, within the type 2 family of modes, one can distinguish a few patches (four in this case) of modes (see rectangles in inset in Fig.~\ref{fig:Eigenvalues}), which we denote as \textit{polaritonic} modes for reasons explained below. More precisely, for a single realization, the spectrum of eigenfrequencies always contains the same number of polaritonic modes, with each mode falling into one of these patches~\footnote{Due to the symmetry of the average atomic sample, polaritonic modes are actually found in pairs with very close eigenfrequencies for each realization.}. Figure~\ref{fig:Excitation}a shows a plot of the dipole moments squared $|\textbf{p}_j|^2$, for a typical mode of type 2. For this collective mode, most of the atoms are excited and Fig.~\ref{fig:Excitation}a suggests  $|\textbf{p}_j|^2$ oscillates spatially along the $z$-axis. We have checked that this oscillation is an intrinsic property of the homogeneous system, i.e. that it is a property of the shape and density of the sample but not of the exact positions of the atoms. To do this, we calculate the average dipole moment squared $\langle|\textbf{p}|^2\rangle$, where $\langle\cdot\rangle$ denotes averaging both inside thin slices perpendicular to the long axis of the box, and over $300$ realizations of the uniform atomic distribution. Only the modes inside the black rectangle in inset of Fig.~\ref{fig:Eigenvalues} are taken into account in the average. Figure~\ref{fig:Excitation}b shows that the spatial oscillation observed in Fig.~\ref{fig:Excitation}a survives the ensemble averaging. Note that the symmetry of the macroscopic mode structure with respect to the plane $z=0$ is restored by the averaging procedure. These facts confirm that these few type 2 eigenmodes are polaritonic by nature. This is in contrast with the other types of modes, which do not show this spatial periodicity along the $z$-axis nor involve all the dipoles simultaneously (see~\cite{SM} for examples of modes of types 1 and 3).

We now study light scattering by the cloud of atoms.
Solving the coupled-dipoles equations using the laser field as an external driving source allows to calculate the scattered field $\textbf{E}$ in the far field for a given realization of the cloud when it is illuminated by an $x$-polarized plane wave, as is done in our recent work~\cite{Stephan}. After averaging over $1500$ realizations, the scattering pattern obtained near resonance ($\omega=\omega_0$) exhibits one lobe in the forward direction (see Fig.~\ref{fig:Sample}a) and some diffuse light which is on average quasi-isotropic and has a smaller amplitude. However, these two features have comparable contributions when integrated over the full solid angle with $54\%$ ($46\%$) of the total scattered light, respectively. The lobe in the forward direction is very similar to the diffraction pattern originating from a homogeneous macroscopic object, suggesting that one could replace the cloud, with its random graininess, by a homogeneous cloud with the same shape and extract an effective permittivity.

We understand these observations by decomposing the electric field $\textbf{E}$ scattered by the random medium into an ensemble-averaged field $\langle \textbf{E}\rangle$ (also denoted coherent) and a fluctuating field $\delta\textbf{E}$ (also denoted incoherent), $\textbf{E}=\langle\textbf{E}\rangle+\delta\textbf{E}$, with $\langle\delta\textbf{E}\rangle=0$~\cite{Frisch,Lagendijk}. $\langle \textbf{E}\rangle$ satisfies the Helmholtz equation in an effective homogeneous medium with a dielectric permittivity $\varepsilon_\text{eff}$ : $\nabla^2\langle\textbf{E}\rangle+\frac{\omega^2}{c^2}\varepsilon_\text{eff}\langle\textbf{E}\rangle=0$~\cite{Frisch,Lagendijk}.
From this it follows that $\langle \textbf{E}\rangle$ is the field diffracted by an effective medium with dielectric permittivity $\varepsilon_\text{eff}$ and with the same shape as the cloud. The scattered intensity can also be decomposed in a coherent and incoherent contribution $\langle|\textbf{E}|^2\rangle=|\langle\textbf{E}\rangle|^2+\langle|\delta\textbf{E}|^2\rangle\propto I_{\text{coh}}+I_{\text{incoh}}$. The first term corresponds to the lobe in the forward direction in Fig.~\ref{fig:Sample}a, while the second term is the quasi-isotropic diffuse light.

\begin{figure}
\includegraphics[width=\linewidth]{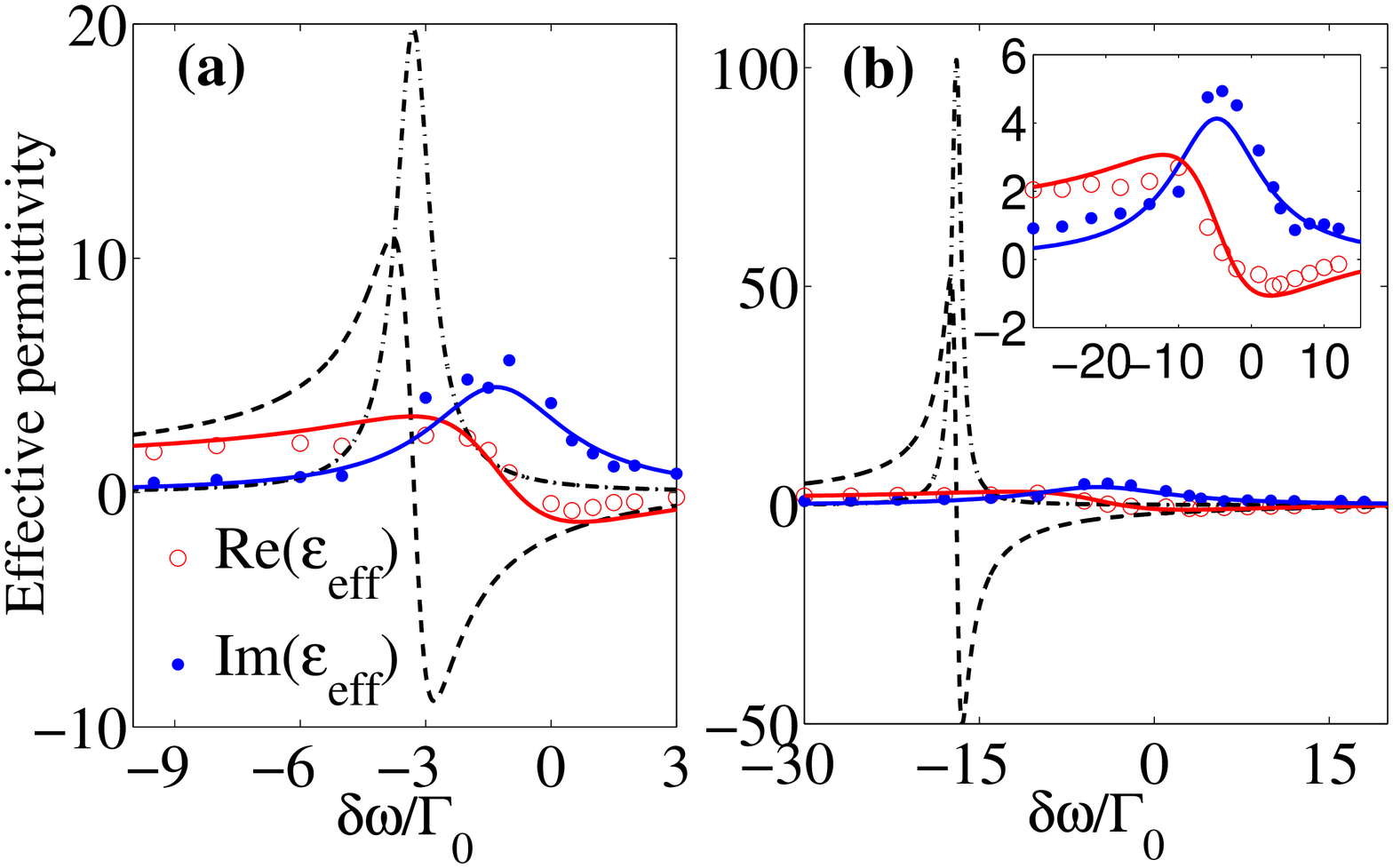}
\caption{Effective permittivity $\varepsilon_\text{eff}(\omega)$ for (a) $\rho/k^3=1$, and (b) $\rho/k^3=5$. The real and imaginary parts of the effective permittivity (circles and dots) largely differ from the predictions of Lorentz-Lorenz theory (dashed curves). Solid lines : Lorentzian fit of the numerical data.}
\label{fig:Dispersion1}
\end{figure}

We now proceed to the extraction of the effective permittivity of the cloud as a function of frequency. Textbook mean-field theory gives for low-dense media the Lorentz-Lorenz relation between the macroscopic permittivity and the polarizability of the scatterers
\begin{equation}\label{eq:CM}
\frac{\varepsilon_{\text{LL}}(\omega)-1}{\varepsilon_{\text{LL}}(\omega)+2}=\frac{\rho\alpha(\omega)}{3}.
\end{equation}
This formula takes only partially dipole-dipole interactions into account by the local-field correction~\cite{SolidState}. Recurrent scattering is not included, while its impact on the effective properties is expected to be of importance for dense systems~\cite{Chomaz,Lagendijk,Janne2014}. In order to derive the effective dielectric constant by accounting for all multiple scattering effects, we solve an inverse problem. We compare the coherent contribution $|\langle\textbf{E}\rangle|^2$ of the far-field scattering pattern of the atomic cloud with that of an effective homogeneous particle with a permittivity $\varepsilon_\text{eff}$ and the same geometry as the cloud. The latter is numerically calculated with an aperiodic Fourier modal method (a-FMM)~\cite{JPH1}, using $\varepsilon_{\text{eff}}$ as a fitting parameter. We find that the coherent far-field scattering pattern agrees with the microscopic pattern within $1\%$ (compare Figs.~\ref{fig:Sample}a and b) showing that the effective permittivity approach is valid. We conclude that this macroscopic approach captures both the recurrent scattering and the collective effects of the microscopic picture. By repeating this procedure for different frequency detunings $\delta\omega=\omega-\omega_0$, we obtain the spectrum of the effective permittivity (see Fig.~\ref{fig:Dispersion1}), which is found to be significantly different from the prediction by Eq.~(\ref{eq:CM}).
Our results evidence that the Lorentz-Lorenz formula is not valid for a dense cloud of resonant scatterers; the discrepancy between the simulation and the Lorentz-Lorenz theory increases with the density because dipole-dipole interactions become stronger. This result has been recently confirmed experimentally using a cloud of cold atoms with similar shape and density~\cite{Stephan}. It is all the more remarkable as the system is five orders of magnitude more dilute than a gas at ambiant conditions for which the Lorentz-Lorenz theory is an accurate description.

It is noteworthy that at resonance the gas is described by a permittivity with a negative real part and thus optically behaves as a metal. From the permittivity we derive the imaginary part of the refractive index $n_\text{eff}=\varepsilon_\text{eff}^{1/2}$ and thus find the mean free path at resonance, $l=1/[k \text{Im}(n_{\text{eff}})]$. We find $l=1/(1.5k)$, indicating that the Ioffe-Regel criterion is satisfied~\cite{Sheng}. Yet, we have found many modes extended throughout the whole system, indicating that the localization length, if any, is larger than the size of the system.

We have seen that the far-field scattering of the atomic cloud is very similar to that of a homogeneous cloud with an effective permittivity. We now make explicit the relation between the (macroscopic) modes of the homogeneous cloud and the (microscopic) polaritonic modes. To do that, we use the fitted effective permittivity $\varepsilon_\text{eff}(\omega)$ (see solid lines in Fig.~\ref{fig:Dispersion1}) and calculate the macroscopic modes of the homogeneous cloud, which are poles of the scattering matrix, by iteratively solving Maxwell's equations in the complex frequency plane~\cite{JPH2}. Remarkably, we find that the macroscopic modes coincide (within error bars) with the above-mentioned polaritonic modes (see Fig.~\ref{fig:Eigenvalues}). Despite the fact that the geometrical length of the cloud is fixed and the frequency of the laser is almost fixed (close to $\omega_0$), it is possible to find several longitudinal modes because of the strong dispersion of the medium close to resonance. This provides a physical explanation of the results reported by Li \textit{et al.}~\cite{Evers}, who studied the spontaneous emission spectrum of a dense cloud of atoms and found blue and red-shifted modes.

Figure~\ref{fig:Excitation} shows that the above-mentioned coincidence is not accidental: the microscopic and macroscopic modes have not only the same frequency and linewidth, but also the same spatial structure along the $z$-axis. This is true also transversally~\cite{SM}. Also, this coincidence is not restricted to the interaction strength parameter studied here (see~\cite{SM} for a study of the modes and scattering pattern of a similar system with $\rho/k^3=5$.). Our analysis thus sheds new light on the connection between the microscopic and macroscopic approaches of scattering, as one can directly identify the (microscopic) polaritonic modes with the macroscopic modes of the corresponding effective homogeneous object.

In the macroscopic model, the appearance of an imaginary part of the effective permittivity (see Fig.~\ref{fig:Dispersion1}) accounts for losses, which correspond to a radiative transfer of energy from the coherent field, along its propagation through the sample, to the incoherent (diffuse) field. While losses are generally considered as being irreversible as a result of dephasing processes, e.g. coupling to phonons, we note that this cannot be the case here, as there is no loss mechanism in the microscopic model. Before averaging, the field scattered by a single realization of the cloud can be time-reversed. After averaging only the coherent field has a well-defined phase and can be time-reversed. In summary, in the presence of dephasing processes, or when the positions of the atoms are changed randomly from one realization to another (as is the case when we ensemble-average), the coupling of the incident light to dipole fluctuations, leads to an irreversible radiation of incoherent light.

In conclusion, we have shown the existence of polaritonic modes in a dense atomic system. These polaritonic modes do not depend on the atomic positions but only on the shape and size of the cloud and on the atomic density, they are spatially delocalized, and strongly superradiant. We have shown that they can be identified to the macroscopic modes of a homogeneous object with an effective permittivity. These results do not only apply to cold atomic clouds but also to  any dense system of resonant scatterers such as molecules or quantum dots. Our work provides a unified vision of scattering by dense systems of resonant scatterers.

\acknowledgements{
We thank V.~Sandoghdar and R.~Carminati for fruitful discussions.
We acknowledge support from the E.U. through the ERC Starting Grant ARENA, from Triangle de la Physique (COLISCINA project), labex PALM (ECONOMIC project) and Region Ile-de-France (LISCOLEM project).
N.J.~S. is supported by Triangle de la Physique and Universit\'e Paris-Sud.
J.-J.~G. is a senior member of Institut Universitaire de France.\\}

\appendix
\section{Type 1 and type 3 eigenmodes}
In this section, we discuss the properties of type 1 and 3 eigenmodes as defined in Fig.~2 of the main text. In the main text, it has been shown that type 2 eigenmodes have a large collective linewidth. This large collective linewidth can be understood by the periodically arrangement of the dipole moments, which allows phase matching of the radiation by the dipoles and therefore superradiance along the axis of the object. Note that this phase matching only occurs within a rather limited solid angle ($\sim 2\pi/15$). This is why the linewidth is not enhanced by a factor $N$ as it is the case for superradiance by $N$ emitters in a subwavelength volume. As the other modes do not possess such a spatial arrangement, the collective linewidth for both type 1 and 3 is smaller.

Type 1 eigenmodes have either a collective linewidth $\Gamma\approx2\Gamma_0$ or $\Gamma\ll\Gamma_0$ and a large frequency shift. Figure \ref{fig:Examples}a shows a typical eigenmode of type 1. It contains only two excited atoms so that we call it a dimer mode. When two dipoles oscillate in-phase and are very close together, the electric fields of both dipoles interfere constructively. Therefore, the pair of dipoles radiates twice as fast ($\Gamma=2\Gamma_0$) as they would do independently. However, when the dipoles oscillate out of phase, their electric fields interfere destructively, and the pair of dipoles is strongly sub-radiant ($\Gamma\ll\Gamma_0$). The large level splitting is due to the large resonant van der Waals interaction.

All modes we have not discussed sofar belong to type 3. These modes involve either many dipoles distributed throughout the whole volume but without a regular spatial structure (Fig.~\ref{fig:Examples}d) as the one observed in Fig.~3b of the main text, or a localized mode involving only a subset of dipoles (Fig.~\ref{fig:Examples}c). 

In order to demonstrate that neither type 1 nor type 3 modes have a spatial structure, as opposed to polaritonic modes, we have calculated the average dipole moment square along the long axis of the box for these modes. Figure~\ref{fig:Type1} does not possess the oscillatory structure observed for type 2 modes. 

\begin{figure}
\includegraphics[width=\linewidth]{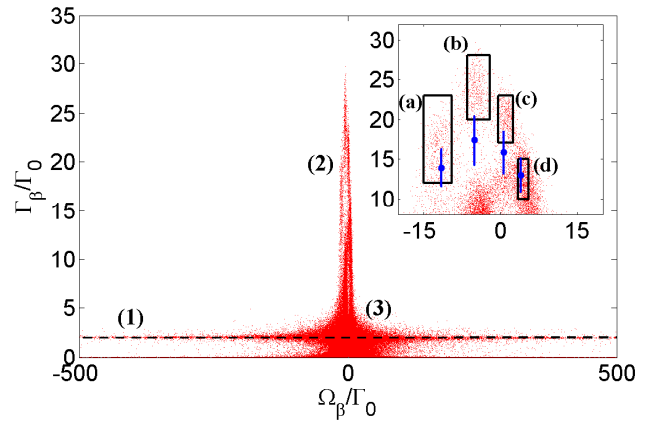}
\caption{The three types of eigenmodes is indicated by the numbers (1) through (3). The inset shows a zoom-in of the main figure, where the polaritonic modes (a), (b), (c) and (d) are visualized in Fig.~\ref{fig:Averaging1}. }
\label{fig:Modes1}
\end{figure}

\begin{figure}
\includegraphics[width=\linewidth]{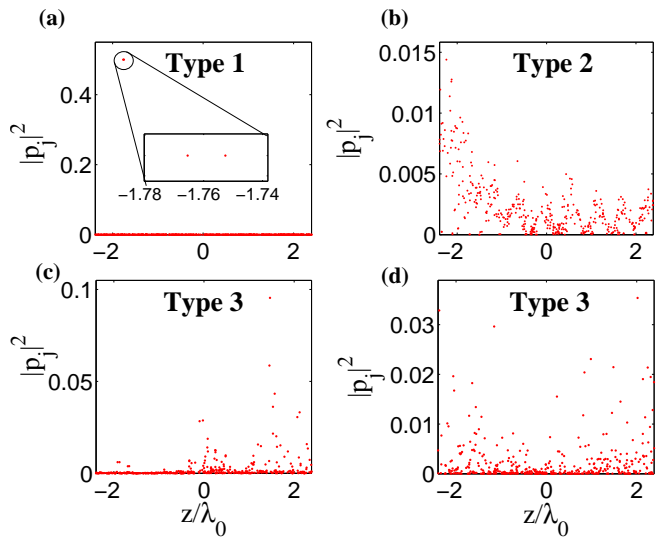}
\caption{Typical eigenmodes are shown. (a) Type 1 eigenmodes are called dimer modes. The dipoles are very close to each other. (b) A polaritonic mode shows an oscillatory behavior. (c) Bunches of dipoles oscillate. (d) Many dipoles oscillate, but without the oscillatory behavior, so there is no periodic arrangement of the dipoles.}
\label{fig:Examples}
\end{figure}

\begin{figure}
\includegraphics[width=\linewidth]{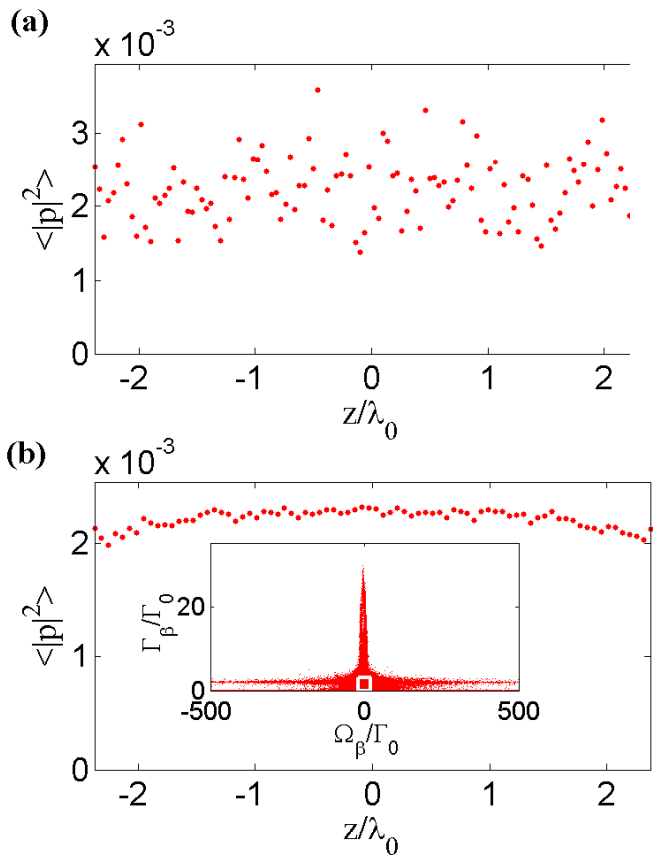}
\caption{The average dipole moment square of type 1 (a) and type 3 (b) eigenmodes along the long axis of the box do not show any spatial structure, like type 2 does. (a) All eigenmodes with $|\Omega_\beta|>150\Gamma_0$ have been taken into account for the averaging. (b) All modes within the white rectangle as depicted in the inset have been taken into account for the averaging.}
\label{fig:Type1}
\end{figure}

\section{Comparing microscopic and macroscopic modes}
As can be seen from Fig. 3 in the main text, the macroscopic mode coincides very well with the microscopic polaritonic mode. In this section, we detail the procedure used to compare the macroscopic mode with the average dipole moment square.

Let us note that the dipole moment square for a single realization (see Fig.~3a of the main text) corresponds to a single eigenmode. In the main text, the procedure of obtaining these eigenmodes has been explained. From the derivation, it follows that these modes are calculated in the \textit{absence} of an external driving field, so the dipole moments are known apart from a multiplication factor. This multiplication factor is obtained by normalizing the modes, which has been done as explained in the main text by imposing $\sum_j |\textbf{p}_j^\beta|^2=1$.

Obviously, the normalization issue also arises for the macroscopic modes. The latter have been normalized such that their intensity coincides well with the average dipole moment square. This way of comparing macroscopic and microscopic modes is consistent with the fact that $\textbf{p}\propto\textbf{E}$. Lastly, we note that $\langle|\textbf{p}|^2\rangle=|\langle\textbf{p}\rangle|^2+\langle|\delta\textbf{p}|^2\rangle$. As a consequence, $\langle|\textbf{p}|^2\rangle$ exhibits an offset, which is due to dipole moment fluctuations. 
This explains the fact that a significant offset as for $\langle|\textbf{p}|^2\rangle$ is not present for $|\textbf{E}|^2$. In order to superimpose them, we have added a constant offset to the macroscopic $|\textbf{E}|^2$.

\section{Polaritonic eigenmodes for $\rho/k^3=1$}
\begin{figure}
\includegraphics[width=\linewidth]{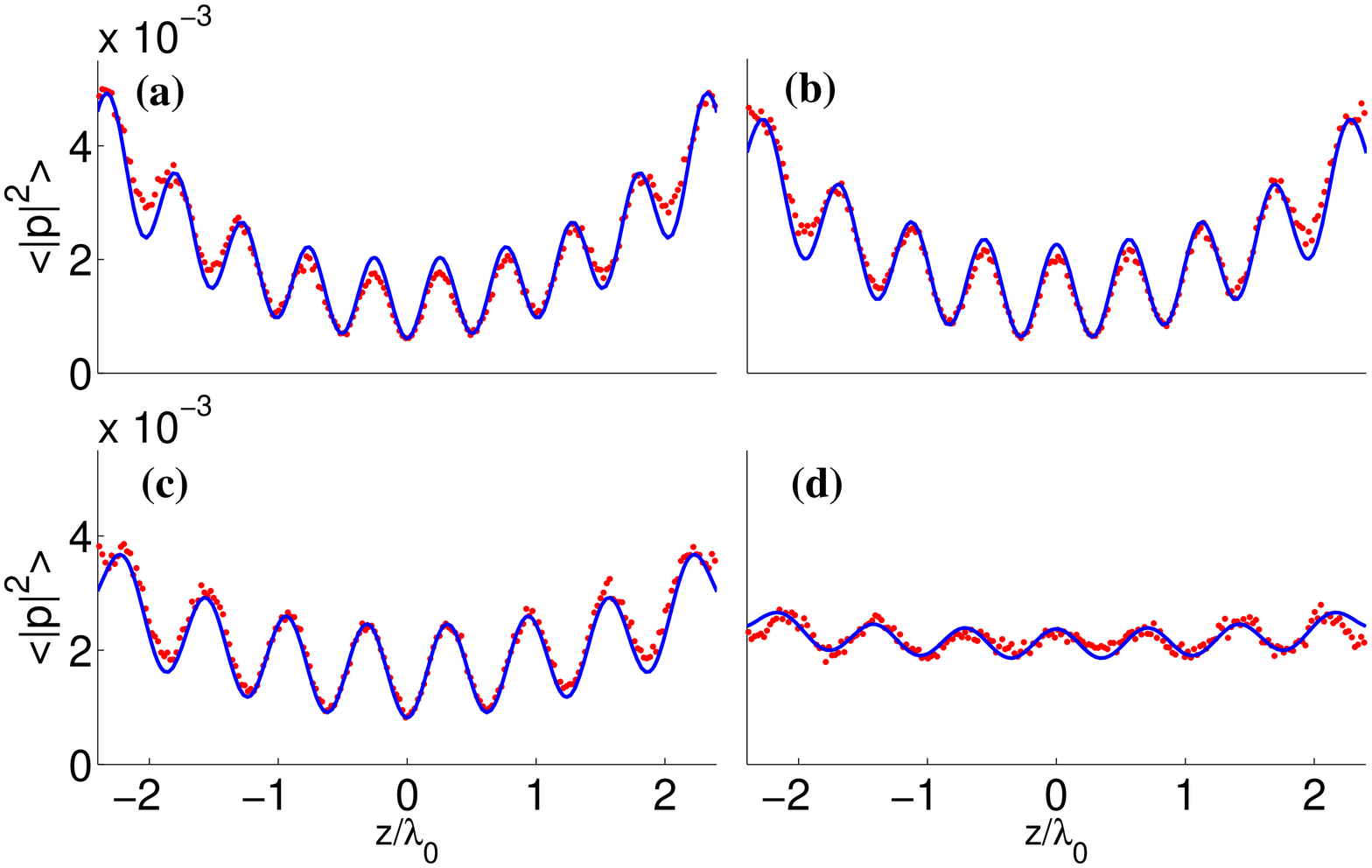}
\caption{The average dipole moment squared of polaritonic (red dots) is plotted along the long axis of the rectangular box. Superimposed (blue solid curve) is $|\textbf{E}|^2$ of the macroscopic mode of the homogeneous object. Each panel belongs to a different polaritonic mode. Refer to Fig.~\ref{fig:Modes1} for the eigenfrequencies of the modes. For the averaging procedure, we divided the system along its long axis in thin slices. The dipole moment squared is averaged per slice and then averaged over 300 realizations.}
\label{fig:Averaging1}
\end{figure}

\begin{figure}
\includegraphics[width=\linewidth]{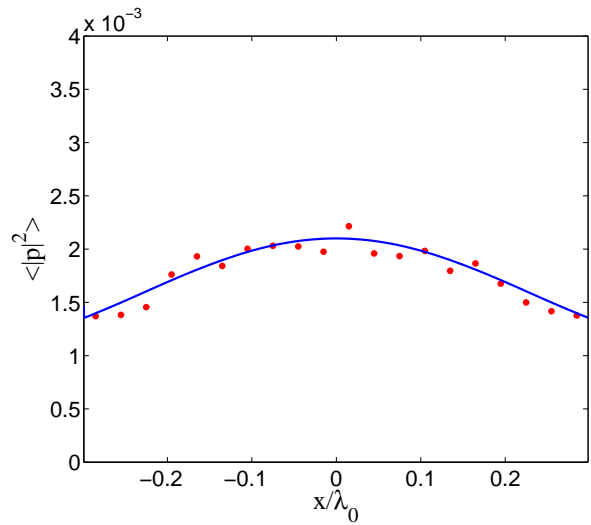}
\caption{Transverse profile of the polaritonic mode (a) of Fig.~\ref{fig:Modes1} (red dots) and of the macroscopic mode (blue solid curves). The profile has been obtained by averaging the dipole moments squared inside a thin slice at $z\simeq -1.3\lambda_0$ and along the $y$ axis.}
\label{fig:ModesTransverse}
\end{figure}

In the main text, we have only shown the spatial distribution of dipole moments for polaritonic mode (a) of Fig.~\ref{fig:Modes1}. For completeness, we show here all polaritonic modes (a) through (d) in Fig.~\ref{fig:Averaging1}. We note that all polaritonic modes can be fitted with the intensity $|\textbf{E}|^2$ of the corresponding macroscopic mode of the homogeneous cloud. Additionally, we observe that the number of oscillations decreases by one from (a) to (d). Modes (a)-(d) are mostly transversally polarized, i.e. in the $(x,y)$-plane. Figure~\ref{fig:ModesTransverse} shows the transverse profile of the polaritonic mode in Fig.~\ref{fig:Averaging1}a.

\section{Eigenmode analysis for $\rho/k^3=5$}
\begin{figure}
\includegraphics[width=\linewidth]{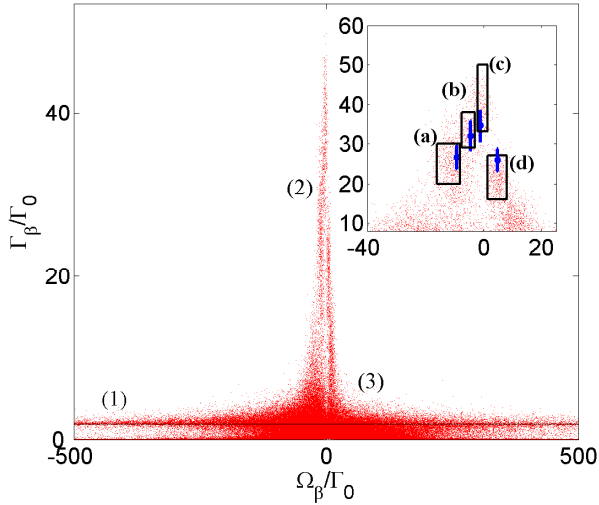}
\caption{The polaritonic modes (a), (b), (c) and (d) are labeled in the same way as for Fig.~\ref{fig:Averaging5}. Inset: enlarged view of the main panel, showing four patches of polaritonic modes (black rectangles). Blue dots: macroscopic modes of the equivalent homogeneous cloud. Error bars are from the Lorentzian fit of the effective permittivity (see main text).}
\label{fig:Modes5}
\end{figure}

\begin{figure}
\includegraphics[width=\linewidth]{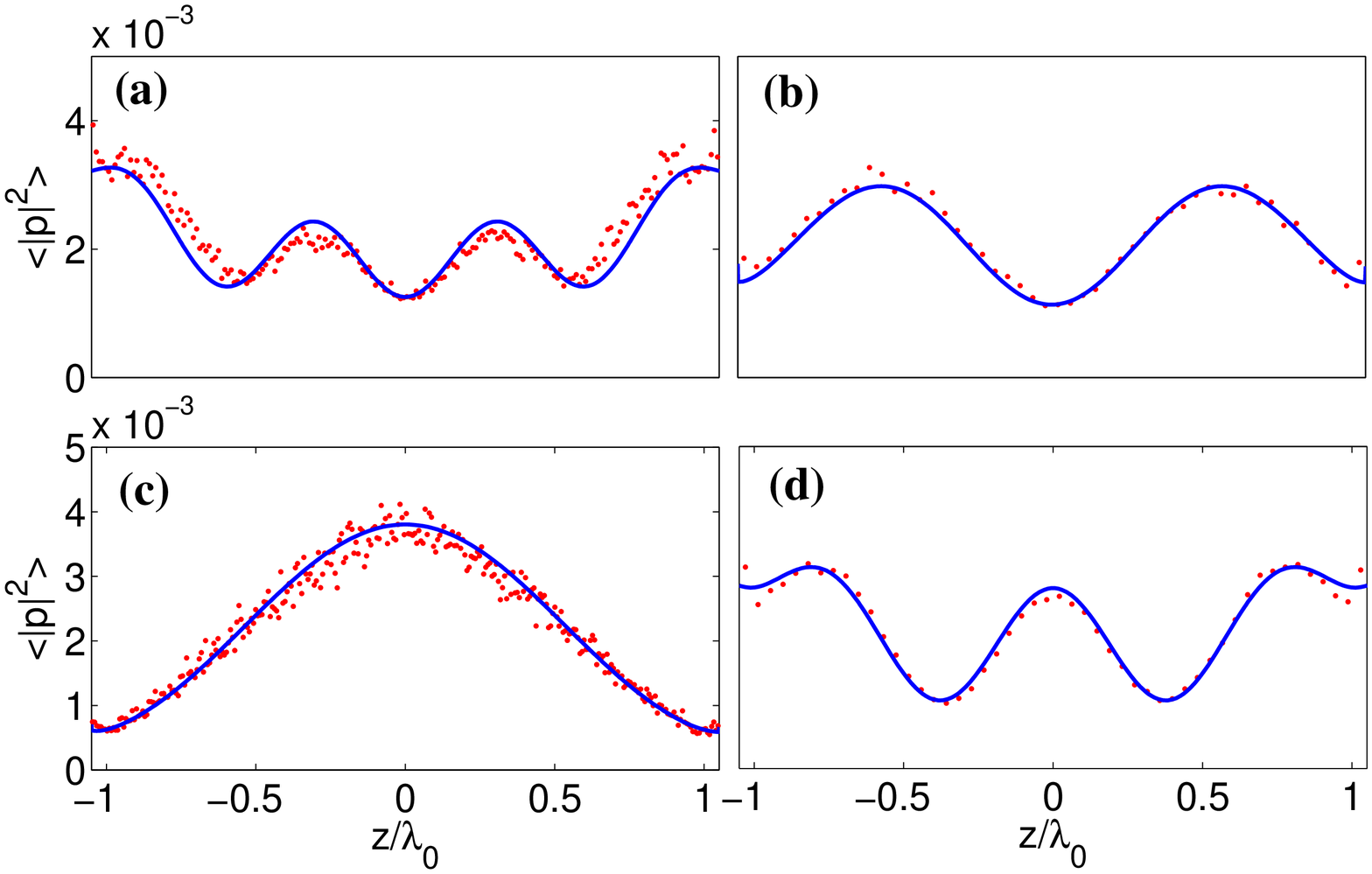}
\caption{The four polaritonic modes are plotted (red dots) with $|\textbf{E}|^2$ of the macroscopic modes superimposed (blue solid curves). The elongated box has been divided in thin slices along the long axis. The dipole moment squared is averaged per slice and then averaged over 300 realizations. }
\label{fig:Averaging5}
\end{figure}

In this section, we show the eigenmodes for the cloud $\rho/k^3=5$, with $N=450$ atoms and dimensions $2.1\lambda_0\times0.4\lambda_0\times0.4\lambda_0$. Figure~\ref{fig:Modes5} shows the distribution of eigenmodes in the complex plane.  We note that also for this density, polaritonic modes are observed. After the averaging procedure, which has been explained in the main text, the four polaritonic modes are visualized in Fig.~\ref{fig:Averaging5}. The modes (a), (b), (c) and (d) correspond to the averaging over the microscopic modes inside the rectangles in the inset of Fig. \ref{fig:Modes5}. Modes (a) and (d) are mostly transversally polarized, i.e. in the $(x,y)$-plane. Modes (b) and (c) are polarized along the long axis of the box ($z$-axis). The same rule as we have seen before holds: the number of oscillations decreases by one when we go to the next mode which has a slightly higher frequency.

\begin{thebibliography}{80}

\bibitem{Bohren}
C.F.~Bohren, and D.R.~Huffman, \textit{Absorption and Scattering of Light by Small Particles} (John Wiley \& Sons, New York, 2008).

\bibitem{Frisch}
U.~Frisch,  in \textit{Probabilistic Methods in Applied Mathematics}, edited by A.A. Bharuch-Reid (Academic, New York, 1968), Vols. I and II.

\bibitem{Sheng}  P.~Sheng, \textit{Introduction to wave scattering, localization, and mesoscopic phenomena}, (Academic Press, New York, 1995).

\bibitem{Lagendijk} A.~Lagendijk, B.~van~Tiggelen, \textit{Resonant multiple scattering of light},
\doilink{10.1016/0370-1573(95)00065-8}{Physics Reports \textbf{270}, 143 (1996)}.

\bibitem{Greffet}
S.~Durant, O.~Calvo-Perez, N.~Vukadinovic, and J.-J.~Greffet, \textit{Light scattering by a random distribution
of particles embedded in absorbing media: diagrammatic expansion of the extinction coefficient}, \doilink{10.1364/JOSAA.24.002943}{J. Opt. Soc. Am. A \textbf{24}, 9 (2007)}.

\bibitem{Sentenac}
P.~Mallet, C.A.~Gu\'{e}rin, and A.~Sentenac, \textit{Maxwell-Garnett mixing rule in the presence of multiple scattering: Derivation and accuracy}, \doilink{10.1103/PhysRevB.72.014205}{Phys. Rev. B \textbf{72}, 014205 (2005)}.

\bibitem{Maxwell} J.C.~Maxwell-Garnett, \textit{Colors in metal glasses and in metallic films}, \doilink{10.1098/rsta.1904.0024}{Phil. Trans. R. Soc. London A  \textbf{203}, 385 (1904)}.

\bibitem{Bruggeman}
D.A.G.~Bruggeman, \textit{Berechnung verschiedener physikalischer Konstanten von heterogenen Substanzen. I. Dielektrizit\"atskonstanten und Leitf\"ahigkeiten der Mischk\"orper aus isotropen Substanzen}, \doilink{10.1002/andp.19354160705}{Ann. Phys. \textbf{24}, 7 (1935)}.

\bibitem{Labeyrie}
G.~Labeyrie, F.~de~Tomasi, J.-C.~Bernard, C.A.~M\"uller, C.~Miniatura, and R.~Kaiser, \textit{Coherent Backscattering of Light by Cold Atoms}, \doilink{10.1103/PhysRevLett.83.5266}{Phys. Rev. Lett. \textbf{83}, 5266 (1999)}.

\bibitem{Wilkowski}
D.~Wilkowski, Y.~Bidel, T.~Chaneli\'ere, D.~Delande, T.~Jonckheere, B.~Klappauf, G.~Labeyrie, Ch.~Miniatura, C.~A.~M\"uller, O.~Sigwarth, R.~Kaiser, \textit{Coherent backscattering of light by resonant atomic dipole transitions}, \doilink{10.1364/JOSAB.21.000183}{J. Opt. Soc. Am. B \textbf{21}, 183 (2004)}.

\bibitem{Labeyrie2} G.~Labeyrie, D.~Delande, C.A.~M\"uller, C.~Miniatura, and R.~Kaiser, \textit{Multiple scattering of light in a resonant medium}, \doilink{10.1016/j.optcom.2004.10.037}{Opt. Commun. \textbf{243}, 157 (2004)}.

\bibitem{Antoine}
J.~Pellegrino, R.~Bourgain, S.~Jennewein, Y.R.P.~Sortais, A.~Browaeys, S.D.~Jenkins, and J.~Ruostekoski, \textit{Observation of Suppression of Light Scattering Induced by Dipole-Dipole Interactions in a Cold-Atom Ensemble}, \doilink{10.1103/PhysRevLett.113.133602}{Phys. Rev. Lett. \textbf{113}, 133602 (2014)}.

\bibitem{Morice}
O.~Morice, Y.~Castin, and J.~Dalibard, \textit{Refractive index of a dilute Bose gas}, \doilink{10.1103/PhysRevA.51.3896}{Phys. Rev. A. \textbf{51}, 3896 (1995)}.

\bibitem{Janne}
 J.~Ruostekoski, and J.~Javanainen, \textit{Quantum field theory of cooperative atom response: Low light intensity}, \doilink{10.1103/PhysRevA.55.513}{Phys. Rev. A \textbf{55}, 513 (1997)}.

\bibitem{Javanainen}
J.~Javanainen, J.~Ruostekoski, Y.~Li, and S.-M.~Yoo, \textit{Shifts of a Resonance Line in a Dense Atomic Sample}, \doilink{10.1103/PhysRevLett.112.113603}{Phys. Rev. Lett. \textbf{112}, 113603 (2014)}.

\bibitem{Havey2011}
Y.A.~Fofanov, A.S.~Kuraptsev, I.M.~Sokolov, and M.D.~Havey, \textit{Dispersion of the dielectric permittivity of dense and cold atomic gases}, \doilink{10.1103/PhysRevA.84.053811}{Phys. Rev. A \textbf{84}, 053811 (2011)}.

\bibitem{Sokolov}
S.~Balik, A.L.~Win, M.D.~Havey, I.M.~Sokolov, and D.V.~Kupriyanov, \textit{Near-resonance light scattering from a high-density ultracold atomic $^{87}$Rb gas}, \doilink{10.1103/PhysRevA.87.053817}{Phys. Rev. A. \textbf{87}, 053817 (2013)}.

\bibitem{Havey3}
K.~Kemp, S.J.~Roof, M.D.~Havey, I.M.~Sokolov, and D.V.~Kupriyanov, \textit{Cooperatively enhanced light transmission in cold atomic matter}, arXiv:1410.2497 (2014).

\bibitem{Chomaz}
L.~Chomaz, L.~Corman, T.~Yefsah, R.~Desbuquois, and J.~Dalibard, \textit{Absorption imaging of a quasi-two-dimensional gas: a multiple scattering analysis}, \doilink{10.1088/1367-2630/14/5/055001}{New Journal of Physics \textbf{14}, 055001 (2012)}.

\bibitem{Akkermans}
L.~Bellando, A.~Gero, E.~Akkermans, and R.~Kaiser, \textit{Cooperative effects and disorder: A scaling analysis of the spectrum of the effective atomic Hamiltonian}, \doilink{10.1103/PhysRevA.90.063822}{Phys. Rev. A \textbf{90}, 063822 (2014)}.

\bibitem{Felinto}
R.A.~de~Oliveira, M.S.~Mendes, W.S.~Martins, P.L.~Saldanha, J.W.R.~Tabosa, and D.~Felinto, \textit{Single-photon superradiance in cold atoms}, \doilink{10.1103/PhysRevA.90.023848}{Phys. Rev. A \textbf{90}, 023848 (2014)}.

\bibitem{Kaiser}
T.~Bienaim\'{e}, R.~Bachelard, N.~Piovella, and R.~Kaiser, \textit{Cooperativity in light scattering by cold atoms}, \doilink{10.1002/prop.201200089}{Fortschr. Phys. \textbf{61}, 2 (2013)}.

\bibitem{Evers}
Y.~Li, J.~Evers, W.~Feng, S.-Y.~Zhu,  \textit{Spectrum of collective spontaneous emission beyond the rotating-wave approximation}, \doilink{10.1103/PhysRevA.87.053837}{Phys. Rev. A \textbf{87}, 053837 (2013)}.

\bibitem{Havey1}
I.M.~Sokolov, M.D.~Kupriyanova, D.V.~Kupriyanov, and M.D.~Havey, \textit{Light scattering from a dense and ultracold atomic gas}, \doilink{10.1103/PhysRevA.79.053405}{Phys. Rev. A \textbf{79}, 053405 (2009)}.

\bibitem{Hopfield1958}
J.J.~Hopfield, \textit{Theory of the contribution of excitons to the complex dielectric constant of crystals}, \doilink{10.1103/PhysRev.112.1555}{Phys. Rev. \textbf{112}, 1555 (1958).}

\bibitem{Bellessa} J.~Bellessa, C.~Bonnand, J.C.~Plenet, J.~Mugnier, \textit{Strong Coupling between Surface Plasmons and Excitons in an Organic Semiconductor}, \doilink{10.1103/PhysRevLett.93.036404}{Phys. Rev. Lett. \textbf{93}, 036404 (2004)}.

\bibitem{Ebbesen} J.~Dintinger, S.~Klein, F.~Bustos, W.L.~Barnes, and T.W.~Ebbesen, \textit{Strong coupling between surface plasmon-polaritons and organic molecules in subwavelength hole arrays}, \doilink{10.1103/PhysRevB.71.035424}{Phys. Rev. B\textbf{71}, 035424 (2005)}.

\bibitem{Torma} T.K.~Hakala, J.J.~Toppari, A.~Kuzyk, M.~Pettersson, H.~Tikkanen, H.~Kunttu, and P.~T\"orm\"a, \textit{Vacuum rabi Splitting and Strong-Coupling Dynamics for Surface-Plasmon Polaritons and Rhodamine 6G Molecules}, \doilink{10.1103/PhysRevLett.103.053602}{Phys. Rev. Lett. \textbf{103}, 053602 (2009)}.

\bibitem{Bellessa2} S.~Aberra~Guebrou, C.~Symonds, E.~Homeyer, J.C.~Plenet, Yu.~N.~Gartstein, V.M.~Agranovich, and J.~Bellessa, \textit{Coherent Emission from a Disordered Organic Semiconductor Induced by Strong Coupling with Surface Plasmons}, \doilink{10.1103/PhysRevLett.108.066401}{Phys. Rev. Lett. \textbf{108}, 066401 (2012)}.

\bibitem{Torma2} L.~Shi, T.K.~Hakala, H.T.~Rekola, J.-P.~Martikainen, R.J.~Moerland, and P.~T\"orm\"a, \textit{Spatial Coherence Properties of Organic Molecules Coupled to Plasmonic Surface Lattice Resonances in the Weak and Strong Coupling Regimes}, \doilink{10.1103/PhysRevLett.112.153002}{Phys. Rev. Lett. \textbf{112}, 153002 (2014)}.

\bibitem{Wiersma90} H.~Fidder, J.~Knoester and D.A.~Wiersma, \textit{Superradiant emission and optical dephasing in J-aggregates}, \doilink{10.1016/0009-2614(90)85258-E}{Chem. Phys. Lett. \textbf{171}, 529 (1990)}.

\bibitem{Scully}
A.A.~Svidzinsky, J.-T.~Chang, and M.O.~Scully, \textit{Cooperative spontaneous emission of $N$ atoms: Many-body eigenstates, the effect of virtual Lamb shift processes, and analogy with radiation of $N$ classical oscillators}, \doilink{10.1103/PhysRevA.81.053821}{Phys. Rev. A \textbf{81}, 053821 (2010)}.

\bibitem{Novotny2006} L.~Novotny, and B.~Hecht, in \textit{Principles of Nano-Optics}, (Cambridge University Press, Cambridge, 2006).

\bibitem{SolidState}
N.W.~Ashcroft, and N.D.~Mermin, in \textit{Solid State Physics}, (Holt, Reinhart, and Winston, New York, 1976).

\bibitem{Janne2014}
J.~Javanainen, and J.~Ruostekoski, \textit{Strong correlations in light propagation beyond the mean-field theory of optics}, arXiv:1409.4598 (2014).

\bibitem{JPH1}
E.~Silberstein, P.~Lalanne, J.-P.~Hugonin, and Q.~Cao, \textit{Use of grating theories in integrated optics}, \doilink{10.1364/JOSAA.18.002865}{J. Opt. Soc. Am. A, \textbf{18}, 2865 (2001)}.

\bibitem{Stephan}
S.~Jennewein, M.~Besbes, N.J.~Schilder, S.D.~Jenkins, C.~Sauvan, J.~Ruostekoski, J.-J.~Greffet, Y.R.P.~Sortais, and A.~Browaeys, \textit{Observation of the failure of Lorentz local field theory in the optical response of dense and cold atomic systems}, arXiv:1510.08041 (2015).

\bibitem{JPH2}
Q.~Bai, M.~Perrin, C.~Sauvan, J.-P.~Hugonin, and P.~Lalanne, \textit{Efficient and intuitive method for the analysis of light scattering by a resonant nanostructure}, \doilink{10.1364/OE.21.027371}{Opt. Express, \textbf{21}, 27371 (2013)}.

\bibitem{SM} Supplementary Material.

\end{thebibliography}
\end{document}